\documentclass[preprint2]{aastex}

\usepackage{graphicx}	
\usepackage{amsmath}	
\usepackage{amssymb}	
\usepackage{multicol}        
\usepackage{bm}		
\usepackage{pdflscape}	

\usepackage[T1]{fontenc}
\usepackage{ae,aecompl}

\begin{document}


\title{Early Science with the Large Millimeter Telescope: an energy-driven wind revealed by massive molecular and fast X-ray outflows in the Seyfert Galaxy IRAS~17020+4544}

\author{A.L. Longinotti\altaffilmark{1,14}, O. Vega\altaffilmark{1}, Y. Krongold\altaffilmark{2}, I. Aretxaga\altaffilmark{1}, M.~Yun\altaffilmark{3},V.~Chavushyan\altaffilmark{1}, C. Feruglio\altaffilmark{4}, A. Gomez-Ruiz\altaffilmark{1,14}, A. Monta\~na\altaffilmark{1,14}, J. Le\'on-Tavares\altaffilmark{5}, A. Olgu\'in-Iglesias\altaffilmark{1},  M. Giroletti\altaffilmark{6}, M. Guainazzi\altaffilmark{7},  J. Kotilainen\altaffilmark{8}, F. Panessa\altaffilmark{9}, 
L. A. Zapata\altaffilmark{10}, I. Cruz-Gonzalez\altaffilmark{2}, V.M. Pati\~no-\'Alvarez\altaffilmark{11}, D.~Rosa-Gonzalez\altaffilmark{1}, A. Carrami\~nana\altaffilmark{1}, L. Carrasco\altaffilmark{1}, E. Costantini\altaffilmark{12}, D. Dultzin\altaffilmark{2}, J.~Guichard\altaffilmark{1}, I. Puerari\altaffilmark{1} and M. Santos-Lleo\altaffilmark{13}}

\affil{1 Instituto Nacional de Astrof\'isica, \'Optica y Electr\'onica, Luis E. Erro 1, Tonantzintla, Puebla, M\'exico, C.P. 72840}
\affil{2 Instituto de Astronomia, Universidad Nacional Autonoma de M\'exico,  Circuito Exterior, Ciudad Universitaria, Ciudad de M\'exico 04510, M\'exico} 
\affil{3 Department of Astronomy, University of Massachusetts, MA 01003, USA} 
\affil{4 INAF Osservatorio Astronomico di Trieste, Via G. Tiepolo 11, Trieste, Italy}
\affil{5 Centre for Remote Sensing and Earth Observation Processes (TAP), Flemish Institute for Technological Research (VITO), Boeretang 282, B-2400 Mol, Belgium}
\affil{6 INAF Istituto di Radioastronomia, via Gobetti 101, 40129, Bologna Italy }
\affil{7 European Space Agency, European Space Research \& Technology Centre (ESTEC), Postbus 299, 2200 AG Noordwijk, The
Netherlands}
\affil{8 FINCA, Dept. of Physics and Astronomy, University of Turku, Finland}
\affil{9 Istituto di Astrofisica e Planetologia Spaziali di Roma (IAPS), Via del Fosso del Cavaliere 100, 00133 Roma, Italy }
\affil{10 Instituto de Radioastronom\'\i a y Astrof\'\i sica, Universidad Nacional Aut\'onoma de M\'exico, P.O. Box 3-72, 58090, Morelia, Michoac\'an, M\'exico}
\affil{11 Max-Planck-Institut f\"ur Radioastronomie, Auf dem H\"ugel 69, 53121 Bonn, Germany}
\affil{12 SRON, Netherlands Institute for Space Research, Sorbonnelaan 2, 3584 CA Utrecht, The Netherlands}
\affil{13 European Space Agency - ESAC, P.O. Box, 78 E-28691 Villanueva de la Ca\~nada, Madrid, Spain}
\affil{14 CONACyT-INAOE}


\begin{abstract}
We report on the coexistence of powerful gas outflows observed in millimeter and X-ray data of the Radio-Loud Narrow Line Seyfert 1 Galaxy IRAS~17020+4544. 
Thanks to the large collecting power of the Large Millimeter Telescope, a prominent line arising from the $^{12}$CO(1-0) transition was revealed in recent observations of this source. The complex profile is composed by a narrow double-peak line and a broad wing. While the double-peak structure may be arising in a disk  of molecular material, 
 the broad wing is interpreted as the signature of a massive outflow of molecular gas with an approximate bulk velocity of -660~km~s$^{-1}$. 
This molecular wind is likely associated to  a multi-component X-ray Ultra-Fast Outflow with velocities reaching up to $\sim$~0.1{\it c} and column densities in the range 10$^{21-23.9}$~cm$^{-2}$ that was reported in the source prior to the LMT observations.
The momentum load estimated in the two gas phases indicates that within the observational uncertainties the outflow is consistent with being propagating through the galaxy and sweeping up the gas while conserving its energy. This scenario, which has been often postulated as a viable mechanism of how AGN feedback takes place,  has so far been  observed only in ULIRGs sources. IRAS~17020+4544 with bolometric and infrared luminosity respectively of 5$\times$10$^{44}$ erg~s${-1}$ and 1.05$\times$10$^{11}$$L_{\odot}$ appears to be  an example of AGN feedback in a NLSy1 Galaxy (a low power AGN). New proprietary multi-wavelength data recently obtained on this source will allow us to corroborate the proposed hypothesis. 
\end{abstract}


\section{Introduction}
Feedback from luminous Active Galactic Nuclei (AGN) is widely recognized as a key ingredient for evolution of galaxies (Di Matteo et al. 2005, Hopkins \& Elvis 2010).
The energy released as a result of the accretion of large amount of gas during the earliest stage of a quasar life time acts as a trigger for the ejection of powerful outflows driven by the AGN. The action of such winds would be to sweep the gas and possibly eject it out of the host galaxy, thus providing an effective mechanism of quenching star formation. 

One of the most credited scenarios for explaining how quasar feedback actually works proposes that a sub-relativistic wind  with velocity higher than 10$^{4}$ km~s$^{-1}$  and typically observable in the X-ray band, is launched at accretion disk scale (Faucher-Giguere \& Quataert 2012). This highly ionized X-ray gas is currently observed in the form of Ultra Fast Outflows  in 30-40\% of AGN spectra (Tombesi et al. 2012, Gofford et al. 2013). 

While traveling outward, the wind undergoes a collision with the inter-stellar medium (ISM) that gives rise to shock processes (King 2010).  After shocking with the gas, deceleration and cooling processes take place in the outflow, which keeps moving outward giving rise to less ionized lines,  observable in the optical band (e.g. Harrison et al. 2014, Carniani et al. 2015, Fiore et al. 2017), and to the formation of a bubble of hot tenuous gas (e.g. Zubovas \& King 2012). 
At even larger scales, the effect of the cooling eventually converts the ionized gas to a colder phase medium outflowing at a much lower velocity. This latest phase is frequently observed in several ULIRGs and Quasars (Veilleux et al. 2013, Cicone et al. 2014, Feruglio et al. 2010, 2015).

The outflow properties can be described in term of mass loss rate ($\dot{M}_{{out}}$), outflow velocity (v$_{out}$) and momentum flux ($\dot{M}_{{out}}$$\times$v$_{out}$). The nuclear wind at launching radius carries an amount of energy equal to $\frac{1}{2}$$\dot{M}_{{nucl}}$$\times$v$_{nucl}^2$. If during the expansion the outflow conserves most of its initial energy,  the energy conservation for the large scale outflow leads to $\dot{M}_{{ls}}$$\times$v$_{ls}^2$=$\dot{M}_{{nucl}}$$\times$v$_{nucl}^2$.  The resulting momentum flux of the wind at large scale is then boosted by a factor proportional to the ratio of the outflow velocities $\dot{P}_{{ls}}$=
$\frac{v_{nucl}}{v_{ls}}$ $\times$ $\dot{P}_{{nucl}}$.  These considerations lead to a characteristic prediction for energy conserving outflows that are driven by a nuclear wind whose momentum rate is comparable to the AGN radiative power  $\dot{P}_{{nucl}}$=$\frac{L_{AGN}}{c}$  (see Figure 5 in Faucher-Giguere \& Quataert 2012 or Figure 1 in Nardini \& Zubovas 2018). 

If instead the energy of the nuclear wind is dissipated, the momentum rate of the large scale wind does not receive any  ``boosting factor",  meaning that $\dot{P}_{{ls}}$= $\dot{P}_{{nucl}}$, i.e. the momentum rate at launching radius is equal  to the momentum rate at large scale. In this case, the expansion of the wind follows the prediction of momentum conservation. 

To test and corroborate the fascinating hypothesis of a multi-phase outflow spanning the galaxy at all scales, it is evident that observations have to target as many outflow components as possible in the same object. 
This may be expensive in terms of observational campaigns as, for instances, the innermost X-ray phase of the wind is not easily detectable in the majority of X-ray spectra, or the extended emission of hot gas is too tenuous to be observed in a reasonable amount of telescope time, or the source of interest is at too high redshift to allow fine spatially resolved spectroscopy.  

To date, only two  cases of ULIRGS have been published where the X-ray and molecular phases of the outflow were observed and physically related (Tombesi et al. 2015, Feruglio et al. 2015). Remarkably, both results are in excellent agreement with the prediction of the energy-conserving outflow model outlined above (but see Nardini \& Zubovas 2018 for a recent reassessment of both results). 

In this paper, we report on a multi-phase outflow discovered in the Radio Loud Narrow Line Seyfert 1 Galaxy IRAS~17020+4544 (Wisotzki \& Bade 1997,  Doi et al. 2011), which was observed in the X-rays (Longinotti et al. 2015), in the radio band (Giroletti et al. 2017) and in  molecular gas (this paper). Even at the moderate luminosity of this source (5$\times$10$^{44}$ ergs~s$^{-1}$),  the emerging picture is stunningly coherent with the action of an AGN-driven outflow capable of propagating through the galaxy at all scales.
The cosmological parameters adopted throughout this work are H$_0$=70~km~s$^{-1}$~Mpc$^{-1}$, $\Omega_m$=0.3, $\Omega_{\Lambda}$=0.7. 
At the redshift of the source, {\it z}=0.0604, the spatial scale of 1$\arcsec$ corresponds to a physical scale of 1.1~kpc and the luminosity distance is 250~Mpc. 
\begin{figure} [t]
 \includegraphics[width=1.05\columnwidth]{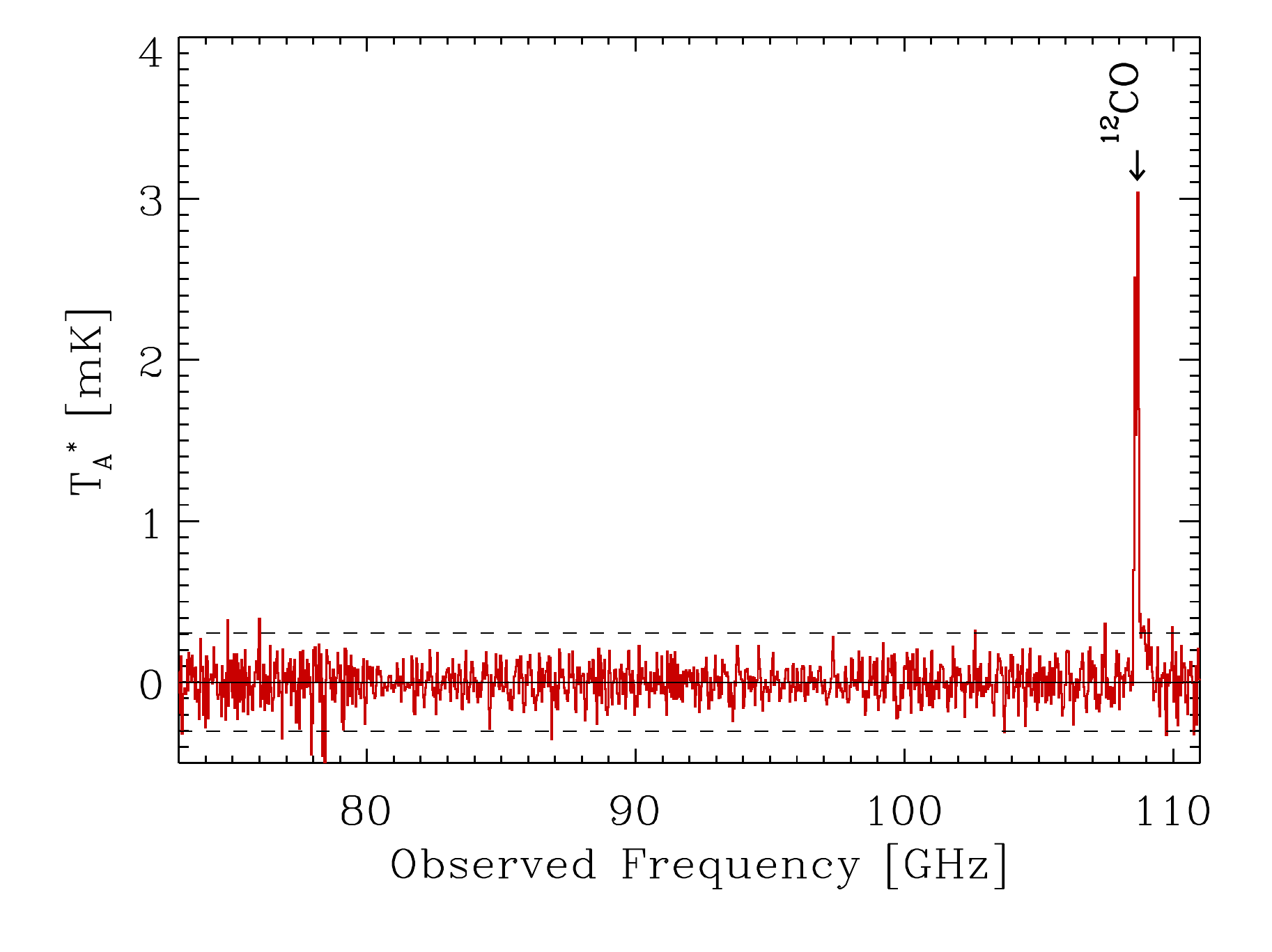}
 \caption{RSR spectrum of the $^{12}$CO(1-0) line in IRAS~17020+4544 obtained with the LMT telescope. The baseline was already subtracted and the horizontal dashed lines mark the $3\sigma$ threshold.  The arrow marks the frequency of the $^{12}$CO(1-0) line at the redshift of IRAS~17020+4544.}
 \label{fig:gtm_spec}
\end{figure}

\section{The data: the Large Millimeter Telescope}

Observations  were obtained at the Large Millimeter Telescope Alfonso Serrano, located on the Sierra Negra Volcano 
in Mexico, at an elevation of 4600 meters (LMT, Hughes et al. 2010) using the  Redshift Search Receiver instrument (RSR, Erickson et al. 2007).  The observing run was carried out in six different nights between May and June 2015, as part of an Early Science program addressed to study the molecular emission of a wide sample of active galactic nuclei (P.I. Le\'on-Tavares). During this phase, only the inner 32.5-m diameter section of the telescope surface was illuminated, leading to an effective beam size of 20~arcsec at 110 GHz.

The RSR has 4 pixels arranged in a dual beam, dual polarization configuration, and the four broad-band receivers cover instantaneously the frequency range 73-111 GHz with a spectral resolution of 31 MHz (R = 3000 or 100 km~s$^{-1}$ at 93 GHz).

IRAS~1702020+4544 was observed for a total of 525 minutes in this configuration over the observing nights, with T$\rm{_{sys}}$ ranging between $\sim$ 100 to 123 K  ($\tau_{225GHz} = 0.2 - 0.4$).  Telescope pointing was checked every 60-90 minutes by observing the source 3C~345. The pointing corrections were always less than 5 arcseconds. The focus was also checked and corrected at the beginning of each run and during sunsets and sunrises.
\begin{figure}[t]
\centering 
 \includegraphics[width=\columnwidth]{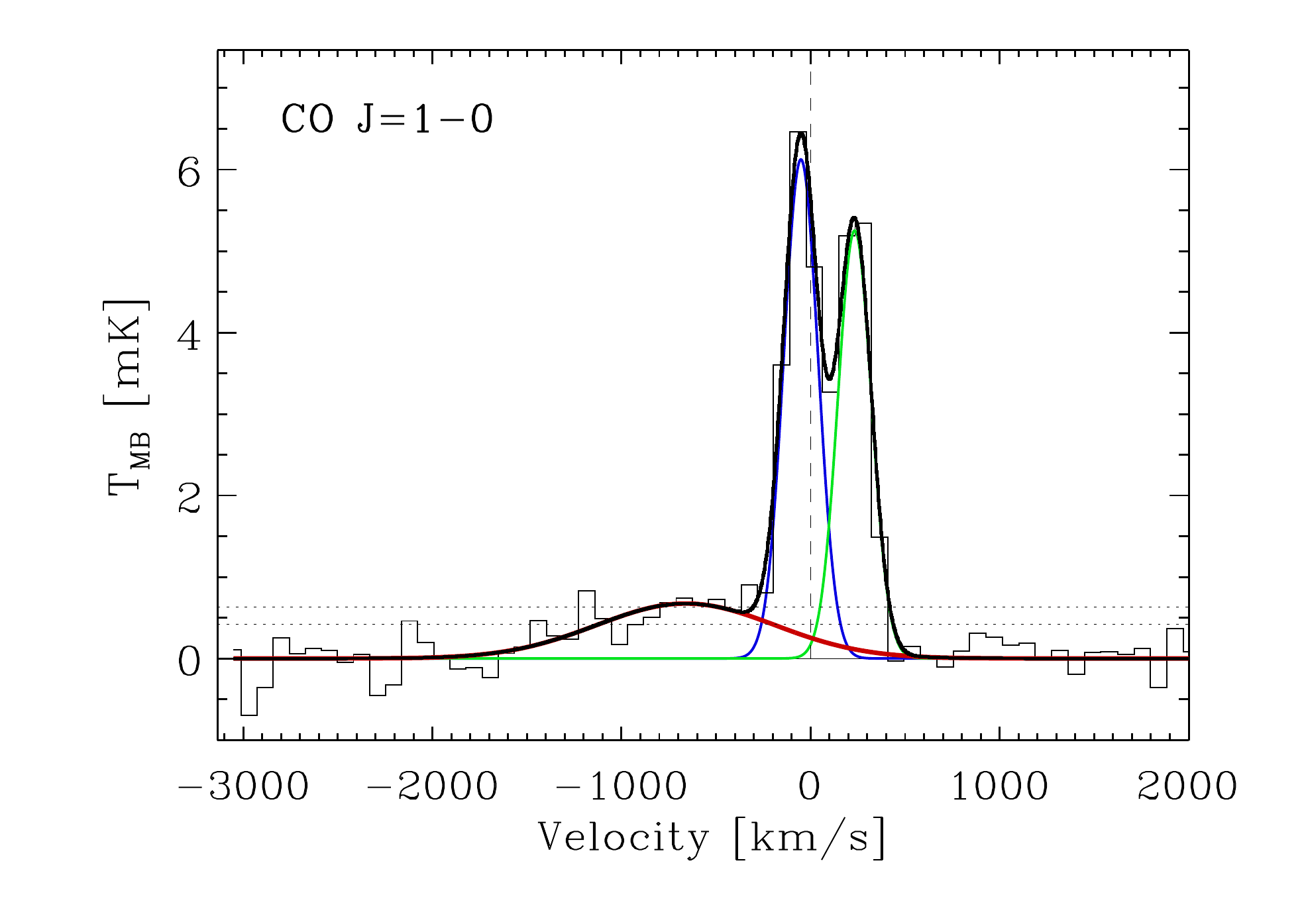}
 \caption{Spectrum of the $^{12}$CO(1-0) line in IRAS~17020+4544
obtained with the LMT telescope. The double-peak structure is fitted by two narrow
Gaussian components (blue and green), and the line wings are
fitted by a very broad Gaussian line (red solid thick line). Dotted horizontal lines mark the 2 and $3\sigma$ thresholds, dashed vertical line marks the zero-velocity position.  Fit parameters are reported in Table 1. }
 \label{fig:gtm_line}
\end{figure}
The observations were calibrated and processed using the latest version
of DREAMPY (Data  REduction  and
Analysis Methods in PYthon), a software package written by G. Narayanan for the purposes of calibrating and reducing LMT - RSR observations. 
Each individual spectrum was analyzed separately. After eliminating bad channels or spectra containing strong ripples, a linear baseline was removed from each spectrum. The
final spectrum was obtained by averaging all
spectra using the $1/\sigma^2$ weight. 
The entire reduced spectrum is obtained in units of
antenna temperatures scale (T$_A^*$), which have been corrected for atmospheric
losses, rear spillover and scattering. We noted that the spectrum still contains some ripples that were not completely removed by our baseline subtraction procedure. Higher order baselines did not improve the resulting spectra. A Savitzky-Golay filter (Savitzky \& Golay 1964) with a second order polynomial and 1 GHz width is applied to the full spectrum to reduce low-frequency residual noise (see also Cybulsky et al. 2016). 
\begin{table*}[t]
\caption{Spectral fit parameters for the CO line complex.}
\footnotesize
\begin{tabular}{ccccccc}
\\
  \hline
Component & FWHM & Centroid & Integrated & L$_{\rm{CO}}$ ($\times$10$^8$) & M$_{\rm{CO}}$ &  $\alpha$   (CO-to-H2)  \\
                    &              &             &   Intensity   &                                                      &                         &      \\ 
& [km~s$^{-1}$] & [km~s$^{-1}$] &  [mK km~s$^{-1}$]  & [K km/s pc$^2$] & [10$^8$ M$_\odot$]  & [M$\odot$ (K~km~s$^{-1}$ pc$^2$)$^{-1}$]  \\
\hline
\\
Broad wing & 1112  & -660 &   798$\pm$252   &   $3.08\pm$ 0.97   & $1.54\pm$ 0.49  &  0.5   \\
Line A         & 213    & -51   &   1390$\pm$114  &   $5.37\pm$ 0.44   & $4.62\pm$ 0.38 &  0.86  \\
Line B         & 210   & 233   &  1171$\pm$110  &   $4.53\pm$ 0.42   & $3.89\pm$ 0.36  &   0.86  \\
\hline
\end{tabular}
\end{table*}

Figure~1 shows the resulting spectrum, which has an overall r.m.s. of 0.1 mK in the 100 GHz spectral region. 
The final spectrum was converted into
main-beam temperatures ($T_{\rm{MB}}$) using the relation $T_{\rm{MB}} =\frac{T_A^*}{\eta_{\rm{MB}}}$, where $\eta_{\rm{MB}}$ is the main-beam efficiency of the telescope, whose value  for that season was  0.5 at 110~GHz.

Only one feature is detected, with a S/N $\sim 28$ at 108.705 GHz, which corresponds to the frequency of $^{12}$CO(1-0) molecular line at the redshift of IRAS~17020+4544. It shows a clear asymmetric double peak structure, with one component peaking at -51 km~s$^{-1}$ and
the other one peaking at +233~km s$^{-1}$ from the zero of the line.

To reproduce the double peak structure, we fitted two narrow Gaussian lines to the profile (see Figure~2). The fit with two Gaussian lines shows residuals in the blue wing significant to $\sim$3$\sigma$.

 The CO intensity of each component is reported in Table~1 together with the corresponding molecular gas mass. For both narrow lines, the molecular gas mass is estimated assuming a CO-to-H2 conversion factor appropriate for ULIRGs (Solomon et al. 1997, Solomon \& Vanden Bout 2005). The molecular gas masses estimated in each narrow component are very similar. This finding is compatible with the presence of a molecular disk of material in the galaxy and will be discussed in a later work (Olgu\'in-Iglesias in prep.). 

Additionally, the data reveals the presence of a broad feature of the $^{12}$CO(1-0) line. We fitted this excess with an additional Gaussian line that is significantly  shifted to the blue with respect to the zero-velocity position. The blue wing extends up to about -1500 km s$^{-1}$ with the bulk of velocity peaking at -660 km s$^{-1}$. The resulting simultaneous fit to the three component is shown in Figure~\ref{fig:gtm_line}, where we also display the integrated intensities, the FWHM, and the corresponding molecular masses of each component. 

The conversion of the CO luminosity of the broad component into molecular
gas mass M(H2) was obtained by assuming a conservative CO-to-H2 conversion factor
$\alpha = 0.5$, i.e. 1/10 the Galactic value (Solomon et al. 1987). This is the lowest conversion factor found in different locations
of M 82 (a typical starburst galaxy), including its molecular outflow (Weiss et al. 2001). 
We note that this value spans a wide range and therefore uncertainties in the estimated mass of the molecular gas can vary accordingly. For example, the CO-to-H2 conversion factor of 2.1 recently estimated by Cicone et al. (2018) in the local merger NGC~6240 is a factor of ~4 higher than the one applied to IRAS17020+4544. However, we decided to adopt the widely used factor of 0.5 in order to ease comparison with previous results (see Section \ref{sec:discussion}).

The broad component is clearly separated from the double peak structure, therefore it can be excluded that this feature is arising in a disk-like geometry as the double peak line. The most viable explanation is that the broad wing of the $^{12}$CO(1-0) line is tracing an outflow of molecular gas whose bulk velocity peaks around -660 km s$^{-1}$.

\subsection{On the significance of the broad wing in the CO line}
Figure~\ref{fig:gtm_spec} shows that the only prominent broad feature in the entire LMT spectrum is the broad blue-shifted component at the base of the double-peaked narrow $^{12}$CO(1-0) line. Although no other broad features are present in the 76-108 GHz and 110-111GHz continua (see Figure~\ref{fig:gtm_spec}), and hence the likelihood of this feature being a spurious detection is negligible, we perform a continuum subtraction simulation to estimate the maximum contribution that any residual baseline could have on the estimated intensity of the broad $^{12}$CO(1-0) line.
 \begin{figure}[t]
 \includegraphics[width=\columnwidth]{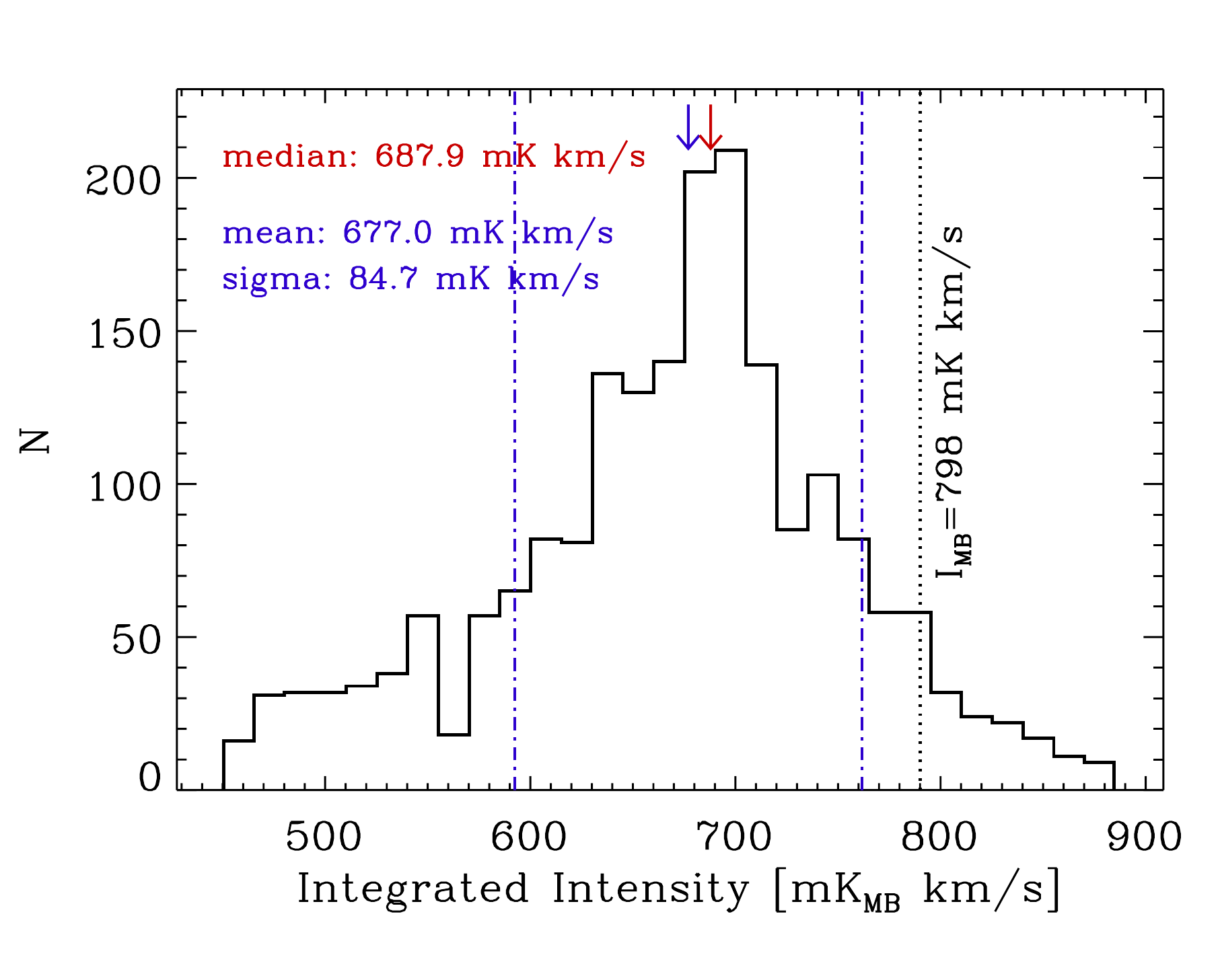}
 \caption{Comparison between the best fit value of the flux in the broad wing of the CO line measured on the LMT spectrum (dotted line) and the results of 2000 trials where the broad line is fitted after randomly subtracting continuum portions of 20 channels extracted from the 76-108 GHz and 110-111GHz continuum interval. }
 \label{fig:montecarlo}
\end{figure}

We choose at random 2000 frequencies in the 76-108 GHz and 110-111GHz range to define windows of 20 channels (i.e. the Full Width at Zero Intensity, FWZI of the feature of interest) that represent plausible fiducial continua  below the detected broad CO line. We subtract these continua and repeat the fitting procedure as above. Figure~\ref{fig:montecarlo} shows the resulting impact on the integrated line flux of the broad component. As expected, there is no chance of the broad component disappearing due to residual bandpasses: for all realizations $L_{\rm CO(broad)}$> 450 mK km~s$^{-1}$. Our originally fitted value of 798~mK km~s$^{-1}$ is never exceeded, and this could be indicative of smaller scale residuals still present in the continuum or the added contribution of faint lines. Our most conservative estimation of the broad line intensity is, hence, given by the median of the distribution in Figure~\ref{fig:montecarlo}:  $L_{\rm CO(broad)}$=690 mK km~s$^{-1}$,  with a 68 per cent confidence interval in the 590-760 mK km~s$^{-1}$. 

We are therefore confident that the detection of the line broad wing is statistically robust and that its astrophysical properties are genuinely derived in this work.

\section{The outflows in IRAS~17020+4544}
\subsection{The molecular outflow}
\label{molout}
The 20~arcsec resolution of the LMT at  110~GHz does not allow us to infer the spatial scale of the CO wind, it only provides an upper limit of 23~kpc at the distance of IRAS~17020+4544.
 Therefore, accounting for the beam size, it is reasonable to say that the present LMT observation can trace molecular gas up to $\sim$20 kpc. Considering that an average AGN lifetime cycle corresponds to 10$^{6-7}$~years, the spatial extension that molecular gas outflowing at the observed velocity can swipe during the AGN life corresponds  to about 0.67-6.7~kpc. These numbers are in excellent agreement also with the dynamical timescale inferred by considering that the molecular gas is moving at a  bulk velocity of -660~km~s$^{-1}$. Therefore, the dynamical timescale T$_{dyn}$ of 10$^{6}$~years for the outflow to propagate out of the nucleus corresponds to a radius R$\sim$0.67~kpc. At this distance, assuming spherical symmetry and that the outflow is moving at the bulk velocity of -660~km~s$^{-1}$, the estimated  mass outflow rate is  $\dot{M}_{[CO]}$=3$\frac{M_{H2} \times v_{out} }{R}$, which yields a mass loss of  $\sim$480~$M_{\odot}$yr$^{-1}$.  If we consider the 
  higher spatial scale of 6.7~kpc, which corresponds to an AGN lifetime of 10$^{7}$~years, the mass outflow rate decreases to 48~$M_{\odot}$yr$^{-1}$.  
These numbers are estimated assuming that the velocity of the outflow is measured as the centroid of the broad line in the LMT spectrum (see Table~1), i.e. as the shift between the peaks of the narrow and broad component. This measurement may be affected by our limited knowledge of the geometry of the outflow, particularly of the inclination of the outflow with respect to our line of sight. 

In this regard, the estimate of the velocity proposed by Rupke \& Veilleux (2013) {\it v$_{out}$}=velocity shift$_{broad}$+2$\sigma$$_{broad}$ where $\sigma$$_{broad}$=$\frac{FWHM_{broad}}{2.35}$,  shall be less affected by the outflow geometry. Hence, we provide the estimate of the mass loss rate for the outflow traveling at v$_{out}$=-1600~km~s$^{-1}$ and  at a distance of 0.67~kpc.  This estimate, which yields $\dot{M}_{[CO]}$=1162~$M_{\odot}$yr$^{-1}$, can be taken as a maximum uncertainty on the amount of mass that the AGN is capable to expel via molecular outflow. All these rates are estimated assuming a uniform distribution of the molecular gas in a spherical geometry with radius R.

The force or momentum flux of this wind is related to the velocity with which the gas is pushed outward i.e. the outflow velocity, therefore it is equivalent to $\dot{P}_{{out}}$=$\dot{M}_{{out}}$$\times$v$_{out}$. Assuming that the gas is outflowing at bulk velocity of -660~km~s$^{-1}$ and assuming that the gas is still confined on the smaller spatial scale of 0.67~kpc, this would yield $\dot{P}_{[CO]}$=$\dot{M}_{[CO]}$$\times$~{\it v}$_{[CO]}$, i.e. $\dot{P}_{[CO]}$$\sim$2$\times$10$^{36}$~cm~g~s$^{-2}$.
Considering the source bolometric luminosity, L$_{bol}$=5$\times$10$^{44}$~erg~s$^{-1}$, the force of the radiation pressure can be expressed as $\dot{P}_{{rad}}$~=~$\frac{L_{bol}}{c}$=1.7$\times$10$^{34}$ cm~g~s$^{-2}$, and conversely,   the ratio of the force of the molecular outflow versus the force of the  radiation would be  $\frac{\dot{P}_{[CO]}}{\dot{P}_{rad}}$ $\sim$117. 
We remark that the bolometric luminosity here considered was estimated accordingly to Longinotti et al. 2015 and by assuming a conservative low bolometric correction ({\it k}=10, see Marconi et al. 2004). 
 If we relax the assumption on the spatial scale maintaining the bulk outflow velocity, we obtained  $\dot{M}_{[CO]}$=48~$M_{\odot}$yr$^{-1}$ and thus $\dot{P}_{[CO]min}$=2$\times$10$^{35}$~cm~g~s$^{-2}$ and $\frac{\dot{P}_{[CO]min}}{\dot{P}_{rad}}$$\sim$11. If we now consider the higher outflow velocity of -1600~km~s$^{-1}$, the resulting momentum flux of the outflow  is boosted to $\dot{P}_{[CO]max}$=1.14$\times$10$^{37}$~cm~g~s$^{-2}$, which leads to a ratio of $\frac{\dot{P}_{[CO]max}}{\dot{P}_{rad}}$$\sim$670 with respect to the momentum of the  AGN radiation.

\subsection{Relation with the X-ray outflow}
The detailed properties of the X-ray wind detected in IRAS~17020+4544 are reported in Longinotti et al. (2015) and Sanfrutos et al. (ApJ accepted). We defer the reader to these publications for all the details on the XMM-Newton data analysis and results. 

Out of the multiple  components that were revealed in the X-ray wind, only one was found sufficiently massive to be able to expel the gas and trigger  negative feedback in the galaxy. The column density and outflow velocity of this X-ray wind  were  N$_{H}$$\sim$10$^{23.9}$~cm$^{-2}$  and v$_{out}$=27,200$\pm$300~km~s$^{-1}$ (Longinotti et al. 2015).
 As outlined in Section~\ref{molout}, the momentum flux of this wind is estimated through the outflow velocity and the mass flux expelled by the wind $\dot{P}_{{X}}$=$\dot{M}_{{X}}$$\times$v$_{out}$. 

The mass and energy outflow rates of the X-ray wind were estimated and parametrized in terms of the wind covering factor: $\dot{M}_{{out}}$$\sim$0.26~C$_f$~$M_{\odot}$yr$^{-1}$ and   $\frac{\dot{E}}{L_{BOL}}$=11\%C$_f$~erg~s$^{-1}$.
The X-ray data presented in Longinotti et al. 2015 did not allow an exact estimate of the geometry and covering factor of the wind. However, we note that the X-ray wind  plausibly originate at accretion disc scale in a conical geometry (e.g. Krongold et al. 2007) and that it has to be sufficiently powerful to push the molecular gas, therefore  its covering factor cannot be much lower than 1. 
We then assume  that C$_f$ may vary in the 0.5-1 range.

Under this assumption, the X-ray wind momentum flux is $\dot{P}_{[X]}$=2.15-4.3$\times$10$^{34}$~cm~g~s$^{-2}$.
Considering the force of the radiation pressure as done in Section~\ref{molout} for the molecular outflow, the ratio of the X-ray wind force versus the radiation force  is estimated $\dot{P}_{[X]}$/$\dot{P}_{rad}$=1.87$\pm$0.62. Remarkably, these error bars are dominated by the uncertainties on the covering factor of the wind rather than by its outflow velocity since the availability of grating spectroscopy (Longinotti et al. 2015) allowed the shift of the absorption lines position to be measured to a higher precision. 

\begin{figure}[t]
 \includegraphics[width=1.\columnwidth]{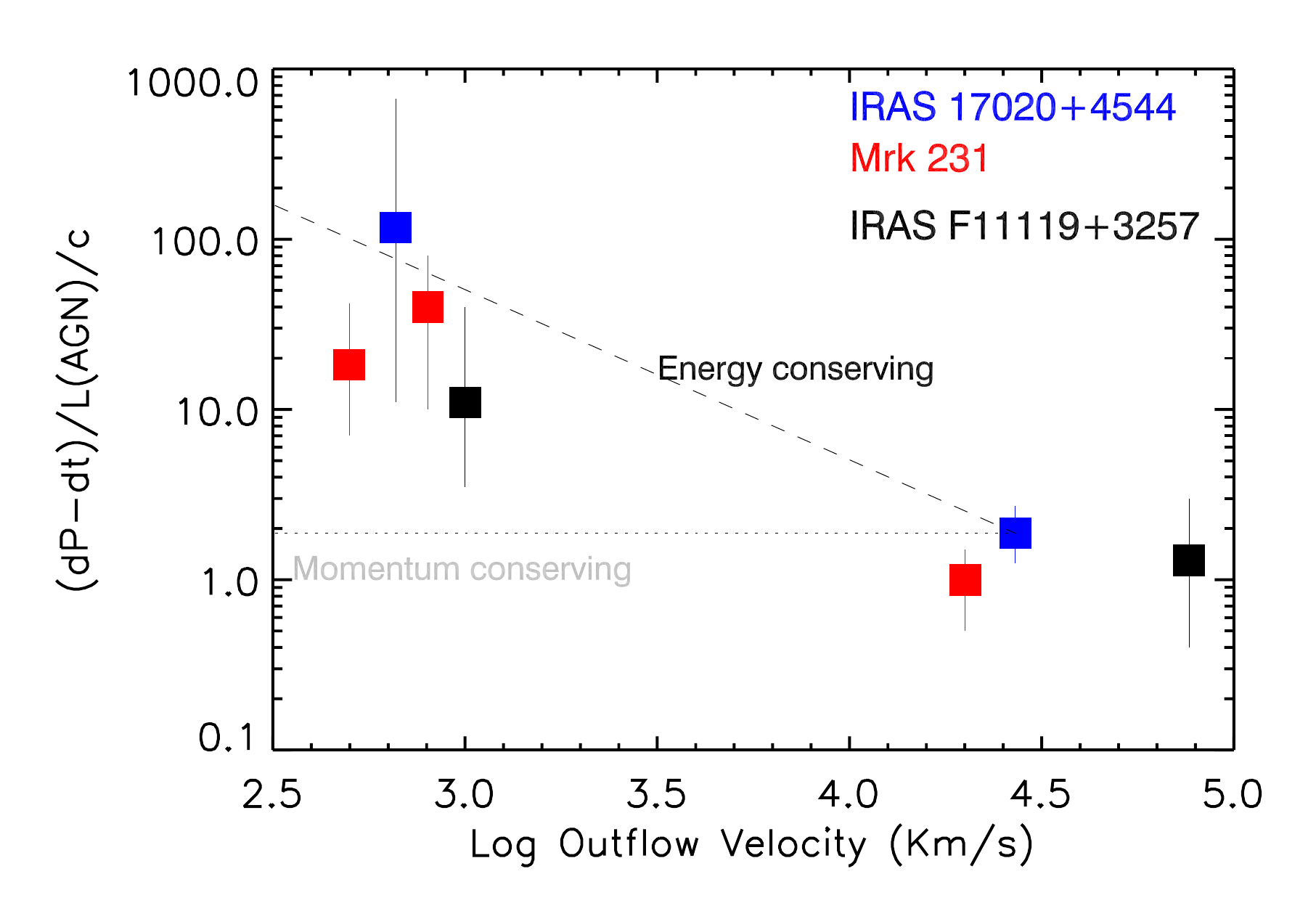}
 \caption{This plot was adapted from Fig.16 in Feruglio et al. 2015 and it shows the force of the molecular and X-ray phases of the wind plotted against their outflow velocities for the three sources where this  relation has been observed. The wind force is expressed in term of the ratio between the flux of the momentum of the wind ($\dot{P}_{[outflow]}$) divided by the force of the radiation $\dot{P}_{rad}$.  The black dashed line marks the prediction for an energy conserving outflow i.e.  
$\dot{P}_{[CO]}$/$\dot{P}_{[X]}$=v$_{out\_X}$/v$_{out\_CO}$ in IRAS~17020+4544. } The grey dotted line marks the prediction for momentum-conserving outflows.
 \label{fig:ppunto}
\end{figure}

\section{Discussion}
\label{sec:discussion}
The results obtained in the two previous sections are summarized in Figure~\ref{fig:ppunto}, which is adapted from Feruglio et al. 2015 and which includes the uncertainties on the outflow distance and the outflow velocity described in Section~\ref{molout}.
This plot compares the prediction for the behaviour of an energy-conserving wind  where the outer molecular outflow is driven by a sub-relativistic wind arising at a much inner scale, as postulated by several authors (Faucher-Giguere \& Quataert 2012, Zubovas \& King 2012 and reference therein).   So far, this theory has been corroborated by observations obtained for the two ULIRGs Mrk 231 (Feruglio et al. 2015) and IRAS~F11119+3257 (Tombesi et al. 2015), which to date provide the only two examples of AGN where an X-ray ultra fast wind and a molecular outflow are observed with momentum loads typical of energy-conserving outflows. 

The result obtained for IRAS~17020+4544 provides a  third indication  for the existence of such energy conserving winds. The uncertainties in the estimates of the mass outflow rates and the wind spatial extent do not allow us to reduce the errors in the calculation of the mutual momentum load of the outflows detected in the X-ray and in the millimeter bands plotted in Figure~\ref{fig:ppunto}. It is anyhow remarkable that the pattern of the outflow force is fairly inconsistent with a wind that conserves its momentum even after accounting for every source of uncertainties. 

This additional piece of evidence for an energy conserving wind that propagates through the galaxy after undergoing a momentum boost, supports the scenario in which powerful AGN outflows are capable of producing feedback on the host galaxy and regulate star formation as proposed in earlier works (Tombesi et al. 2015, and references therein).  

We here anticipate that the outflow properties inferred from the  LMT spectrum are fully supported by proprietary data recently obtained at the NOEMA interferometer that will be presented elsewhere. Interferometry data have been shown to be a powerful tool  to provide refined constraints on the properties of outflowing molecular gas (e.g. Feruglio et al. 2010, 2015), therefore in the future we expect to mitigate the source of uncertainties in the error bars of Figure~\ref{fig:ppunto}  for IRAS17020+4544. 

\subsection{A multi-phase outflow in a Seyfert galaxy at moderate luminosity?}
The case of IRAS~17020+4544 though, bears some important differences from the AGN reported in Figure~\ref{fig:ppunto}. 

The bolometric luminosity of the 2 ULIRGs described in Section~\ref{sec:discussion} is more typical of powerful AGN, being 5$\times$10$^{45}$~erg~s$^{-1}$ for Mrk~231 and  $\sim$~10$^{46}$~erg~s$^{-1}$ for IRAS~F11119+3257 (respectively Feruglio et al. 2015, Tombesi et al. 2015), i.e. at least an order of magnitude larger than in IRAS~17020+4544.
Moreover, as noted already in Longinotti et al. 2015, the infrared luminosity  of L$_{FIR}$= 1.05$\times$10$^{11}$$L_{\odot}$ is not high enough to classify the source as an ULIRG. 

On the contrary, this AGN is hosted by a barred spiral galaxy with no sign of interaction or disturbed morphology (Olgu\'in-Iglesias in prep.), therefore the result here presented  corroborates the evidence that feedback mechanisms typical of gas-rich galaxies at an early evolutionary stage can also arise in lower luminosity spiral galaxies.  

Remarkably, IRAS~17020+4544 shows also peculiar radio properties  (Giroletti et al. 2017) among which an elongated structure produced by synchrotron emission on a 10~pc scale.  This structure is estimated to be moving at a velocity consistent with the X-ray outflows and it might be resolved in a jet in future VLBI observations at higher resolution. 
If confirmed, the presence of such a ``jet" may represent the signature left by the shock of the inner X-ray outflow with the ambient gas that is postulated in several models of galactic outflows (e.g. Zakamska \& Greene 2014; Nims et al. 2015), which attempt to link synchrotron emission in radio-quiet sources with galaxy scale outflows.  

Further hints to the presence of a shocked outflow in this source come from a recent analysis of the connection of the X-ray slow and fast wind (Sanfrutos et al. ApJ accepted, Longinotti 2018), and refined constraints on the energy-conserving scenario are to be expected once IRAS17020+4544 is observed in the UV band by the COS spectrograph onboard HST (PI Y. Krongold) to search for a UV counterpart of the nuclear wind. 

A confirmation of this hypothesis would convert IRAS~17020+4544 in a unique laboratory to study all phases of an energy-driven outflow.

\section*{Acknowledgements}
\addcontentsline{toc}{section}{Acknowledgements}
The authors thank the anonymous referee for the insightful comments and suggestions that have significantly increase the clarity of our manuscript. 
All authors acknowledge technical support from the LMT team at INAOE and at UMASS.   
This work was partially supported by CONACyT research grants 151494 and 280789.  
YK aknowledges support form PAPIIT grant IN06518 from UNAM.
CF acknowledges support from the European Union Horizon 2020 research and innovation program under the Marie Sklodowska-Curie grant agreement No 664931.
JK acknowledges financial support from the Academy of Finland, grant 311438.
LAZ acknowledge financial support from DGAPA, UNAM, and CONACyT, M\'exico. DD acknowledges support through grant IN108716, 
from PAPIIT, UNAM and grant 221398 from CONACyT.  ICG. acknowledges support from DGAPA-UNAM (Mexico) grant IN113417.









\begin{thebibliography}{}
\bibitem[Carniani et al.(2015)]{2015A&A...580A.102C} Carniani, S., Marconi, A., Maiolino, R., et al.\ 2015, \aap, 580, A102
\bibitem[Cicone et al.(2014)]{2014A&A...562A..21C} Cicone, C., Maiolino, R., Sturm, E., et al.\ 2014, \aap, 562, A21
\bibitem[Cicone et al.(2018)]{2018ApJ...863..143C} Cicone, C., Severgnini, P., Papadopoulos, P.~P., et al.\ 2018, \apj, 863, 143
\bibitem[Cybulski et al.(2016)]{2016MNRAS.459.3287C} Cybulski, R., Yun, M.~S., Erickson, N., et al.\ 2016, \mnras, 459, 3287 
\bibitem[Di Matteo et al.(2005)]{2005Natur.433..604D} Di Matteo, T., Springel, V.,  \& Hernquist, L.\ 2005, \nat, 433, 604
\bibitem[Doi et al.(2011)]{2011ApJ...738..126D} Doi, A., Asada, K., \& Nagai, H.\ 2011, \apj, 738, 126
\bibitem[Erickson et al.(2007)]{2007ASPC..375...71E} Erickson, N., Narayanan, G., Goeller, R., \& Grosslein, R.\ 2007, From Z-Machines to ALMA: (Sub)Millimeter Spectroscopy of Galaxies, 375, 71
\bibitem[Faucher-Gigu{\`e}re \& Quataert(2012)]{2012MNRAS.425..605F} Faucher-Gigu{\`e}re, C.-A., \& Quataert, E.\ 2012, \mnras, 425, 605
\bibitem[Feruglio et al.(2010)]{2010A&A...518L.155F} Feruglio, C., Maiolino, R., Piconcelli, E., et al.\ 2010, \aap, 518, L155 
\bibitem[Feruglio et al.(2015)]{2015A&A...583A..99F} Feruglio, C., Fiore, F., Carniani, S., et al.\ 2015, \aap, 583, A99
\bibitem[Fiore et al.(2017)]{2017A&A...601A.143F} Fiore, F., Feruglio, C., Shankar, F., et al.\ 2017, \aap, 601, A143 
\bibitem[Giroletti et al.(2017)]{2017A&A...600A..87G} Giroletti, M., Panessa, F., Longinotti, A.~L., et al.\ 2017, \aap, 600, A87 
\bibitem[Gofford et al.(2013)]{2013MNRAS.430...60G} Gofford, J., Reeves, J.~N., Tombesi, F., et al.\ 2013, \mnras, 430, 60
\bibitem[Harrison et al.(2014)]{2014MNRAS.441.3306H} Harrison, C.~M., Alexander, D.~M., Mullaney, J.~R., \& Swinbank, A.~M.\ 2014, \mnras, 441, 3306 
\bibitem[Hopkins \& Elvis(2010)]{2010MNRAS.401....7H} Hopkins, P.~F., \& Elvis, M.\ 2010, \mnras, 401, 7 
\bibitem[Hughes et al.(2010)]{2010SPIE.7733E..12H} Hughes, D.~H., J{\'a}uregui Correa, J.-C., Schloerb, F.~P., et al.\ 2010, \procspie, 7733, 773312 
\bibitem[King(2010)]{2010MNRAS.402.1516K} King, A.~R.\ 2010, \mnras, 402, 1516
\bibitem[Krongold et al.(2007)]{2007ApJ...659.1022K} Krongold, Y., Nicastro, F., Elvis, M., et al.\ 2007, \apj, 659, 1022
\bibitem[Longinotti et al.(2015)]{2015ApJ...813L..39L} Longinotti, A.~L., Krongold, Y., Guainazzi, M., et al.\ 2015, \apjl, 813, L39 
\bibitem[Longinotti(2018)]{2018arXiv180801043L} Longinotti, A.L.\ 2018, arXiv:1808.01043 
\bibitem{marconi04} Marconi, A., Risaliti, G., Gilli, R., et al.\ 2004, MNRAS, 351, 169
\bibitem[Nardini \& Zubovas(2018)]{2018MNRAS.478.2274N} Nardini, E., \& Zubovas, K.\ 2018, \mnras, 478, 2274
\bibitem[Nims et al.(2015)]{2015MNRAS.447.3612N} Nims, J., Quataert, E., \& Faucher-Gigu{\`e}re, C.-A.\ 2015, \mnras, 447, 3612 
\bibitem[Rupke \& Veilleux(2013)]{2013ApJ...768...75R} Rupke, D.~S.~N., \& Veilleux, S.\ 2013, \apj, 768, 75 
\bibitem[Savitzky \& Golay(1964)]{1964AnaCh..36.1627S} Savitzky, A., \& Golay, M.~J.~E.\ 1964, Analytical Chemistry, 36, 1627 
\bibitem[Solomon et al.(1987)]{1987ApJ...319..730S} Solomon, P.~M., Rivolo, A.~R., Barrett, J., \& Yahil, A.\ 1987, \apj, 319, 730 
\bibitem[Solomon et al.(1997)]{1997ApJ...478..144S} Solomon, P.~M., Downes, D., Radford, S.~J.~E., \& Barrett, J.~W.\ 1997, \apj, 478, 144 
\bibitem[Solomon \& Vanden Bout(2005)]{2005ARA&A..43..677S} Solomon, P.~M., \& Vanden Bout, P.~A.\ 2005, \araa, 43, 677 
\bibitem[Tombesi et al.(2012)]{2012MNRAS.422L...1T} Tombesi, F., Cappi, M., Reeves, J.~N., \& Braito, V.\ 2012, \mnras, 422, L1
\bibitem{Tombesi15} Tombesi, F., Mel{\'e}ndez, M., Veilleux, S., et al.\ 2015, \nat, 519, 436
\bibitem[Veilleux et al.(2013)]{2013ApJ...776...27V} Veilleux, S., Mel{\'e}ndez, M., Sturm, E., et al.\ 2013, \apj, 776, 27 
\bibitem[Zakamska \& Greene(2014)]{2014MNRAS.442..784Z} Zakamska, N.~L., \& Greene, J.~E.\ 2014, \mnras, 442, 784 
\bibitem[Zubovas \& King(2012)]{2012ApJ...745L..34Z} Zubovas, K., \& King, A.\ 2012, \apjl, 745, L34 
\bibitem[Wei{\ss} et al.(2001)]{2001A&A...365..571W} Wei{\ss}, A., Neininger, N., H{\"u}ttemeister, S., \& Klein, U.\ 2001, \aap, 365, 571
\bibitem[Wisotzki \& Bade(1997)]{1997A&A...320..395W} Wisotzki, L., \& Bade, N.\ 1997, \aap, 320, 395


\end{thebibliography}
\end{document}